\documentclass[10pt,technote]{IEEEtran}
\pdfoutput=1
\usepackage{amsmath,amsfonts}
\usepackage{algorithmic}
\usepackage{array}
\usepackage[caption=false,font=normalsize,labelfont=sf,textfont=sf]{subfig}
\usepackage{textcomp}
\usepackage{stfloats}
\usepackage{url}
\usepackage{verbatim}
\usepackage{graphicx}
\def\BibTeX{{\rm B\kern-.05em{\sc i\kern-.025em b}\kern-.08em
    T\kern-.1667em\lower.7ex\hbox{E}\kern-.125emX}}
\usepackage{balance}

\begin{document}

\title{Modeling and Performance Analysis of Single-Server Database Over Quasi-static Rayleigh Fading Channel}

\author{Mengying Chen, Wannian An, Yang Liu, Chen Dong, Xiaodong Xu, \emph{Senior Member, IEEE}, Boxiao Han, Ping Zhang, \emph{Fellow, IEEE}
\thanks{This work was supported in part by State Key Laboratory of Networking and Switching Technology, Beijing University of Posts and Telecommunications, and in part by Beijing University of Posts and Telecommunications-China Mobile Research Institute Joint Innovation Center (Project Number: R207010101125D9) and Fundamental Research Funds for the Central Universities(Project Number: 2021RC01).\emph{(Corresponding author: Chen Dong.)}}
\thanks{Mengying Chen, Wannian An and Yang Liu are with the School of Information and Communication Engineering Beijing University of Posts and Telecommunications(e-mail: chen\_mya@bupt.edu.cn; 1914852893@qq.com; liu\_young@bupt.edu.cn).}
\thanks{Chen Dong, Xiaodong Xu and Ping Zhang are with the State Key Laboratory of Networking and Switching Technology, Beijing University of Posts and Telecommunications, Beijing, China (e-mail:dongchen@bupt.edu.cn; xuxiaodong@bupt.edu.cn; pzhang@bupt.edu.cn).}
\thanks{Boxiao Han is with the Future Mobile Technology Lab China Mobile Research
(e-mail: hanboxiao@chinamobile.com).}}

\markboth{IEEE TRANSACTIONS ON VEHICULAR TECHNOLOGY,~Vol.~X, No.~X, August~202X}%
{Shell \MakeLowercase{\textit{et al.}}: A Sample Article Using IEEEtran.cls for IEEE Journals}


\makeatletter
\def\ps@IEEEtitlepagestyle{%
  \def\@oddfoot{\mycopyrightnotice}%
  \def\@evenfoot{}%
}
\def\mycopyrightnotice{%
  {\hfill \footnotesize 0000--0000/00\$00.00~\copyright~2021 IEEE\hfill}
}
\makeatother
\maketitle
\begin{abstract}
Cloud database is the key technology in cloud computing. The effective and efficient service quality of the cloud database is inseparable from communication technology, just as improving communication quality will reduce the concurrency phenomenon in the ticketing system. In order to visually observe the impact of communication on the cloud database, we propose a Communication-Database (C-D) Model with a single-server database over the quasi-static Rayleigh fading channel, which consists of three parts: CLIENTS SOURCE, COMMUNICATION SYSTEM and DATABASE SYSTEM. This paper uses the queuing model, M/G/1//K, to model the whole system. The C-D Model is analyzed in two cases: nonlinearity and linearity, which correspond to some instances of SISO and MIMO. The program simulation results of average staying time, average number of transactions and other performance characteristics are basically consistent with theoretical results, which verify the validity of the C-D Model. The comparison of these experimental results also proves that poor communication quality does lead to the reduction in the quality of database service.
\end{abstract}


\section{Introduction}
\IEEEPARstart{T}{he} database can store data efficiently and orderly, which makes it convenient for people to use data, such as checking the inventory of train tickets. In order to ensure the stability of online services in the face of a large number of demand, the improvement of the online service system is needed.

As we all know, the online services system mainly consists of communication systems and database systems. The delay of threads based on high concurrency database systems is less if it is in the high quality of communication environment. Therefore, the joint analysis of communication systems and database systems for enhancing the quality of online services is what we consider.

The study in \cite{ulusoy1992simulation} proposed a simulation model within a finite amount of time in a communication network that enables users to analyze distributed transaction scheduling algorithms in real-time database systems. For communication system, most studies assume that the communication network has a constant transmission capacity, such as \cite{garcia1979performance,mariasoosai1990concurrency,singhal1986concurrency}. In \cite{kuang1991performance,ren1996analysis,shyu1990performance}, the models are no longer limited to a constant transmission delay but still assume that the network capacity is infinite. The communication links may become the restriction in some wireless communication systems. With the continuous expansion of communication applications, there are more and more researches on mobile edge
computing(MEC) stack services \cite{du2021resource,du2020mec}. Unfortunately, very few studies have been made to combine a database model with a communication environment.

On the other hand, the performance studies of database systems commonly use queuing systems as the analytical models. Some of the earliest queuing models of database systems can be found in \cite{baccelli1982data} which models the databases through the M/M/m/FCFS queuing system. In \cite{nelson1985analysis}, the model M/M/m is used to study the performance characteristics of a replicated database under synchronous and non-synchronous cases. The studies in \cite{ciciani1990analysis} and \cite{ciciani1992analysis} use M/M/1 queues with an exponential distribution of service times. More generally, \cite{falin1998finite} and \cite{hwang1996data} model the databases as M/G/1/FCFS and M/G/1/RR queues with generally distributed service times, respectively.

All of the preceding articles disregard or make constant communication time. In this paper, we offer a model that combines communication systems with database systems, named the C-D Model, in order to display the status of database service in response to fluctuating channel conditions and arrival rate of transactions.

As a result, a total time distribution for each customer, including communication time and service time, is introduced. Several stationary performance parameters, including queue length (LS, both waiting and serving), waiting queue length (L, only waiting), waiting time (W, only waiting), staying time (WS, both waiting and serving), and busy period of the database (BP), are considered. The main contributions of this paper are summarized as follows:

\emph{1)} A model, named the C-D Model, is designed for describing the impact of communication quality on the services provided by a single-server database.

\emph{2)}  Two simulation outcomes are obtained: linearity and nonlinearity.

\emph{3)} For nonlinear and linear cases, the impact of communication quality is analyzed on a single-server database system, where its program simulation and the theoretical results match each other.

This paper is arranged as follows: The second section describes the C-D Model in detail. In the third section, two cases of unclosed numerical solutions are derived from theoretical analysis. The fourth section explains the program simulation and theoretical results that verify the C-D Model, as well as the effect of communication quality on single-server database service.

\emph{Notation:} The characters in TABLE \ref{tab:table1} are representations of the queuing model. A complete queuing model is expressed as
\begin{equation}
X/Y/Z/A/B/C,
\end{equation}
where the meaning of each letter is as follows: $X$ is the distribution of customer arrival interval, $Y$ is the distribution of service time. The distribution of $X$ and $Y$ is as follows: $M$ represents the negative exponential distribution with a negative parameter; $D$ is deterministic distribution; $E_{k}$ stands for the k-order Erlang distribution, $H_{k}$ is k-phase hyper-exponential distribution, and $G$ represents the general distribution. $Z$ says the number of services. $A$ indicates the system capacity limit, and its default is $+\infty$. $B$ is the limit on the number of client sources, and its default value is $+\infty$. $C$ indicates the service rule, and its default value is the first come, first serve rule.
\begin{table*}[!t]
\begin{center}
\caption{Basic Choices of Modeling\label{tab:table1}}
\centering
\begin{tabular}{|c|c|c|c|c|c|c|}
\hline
Reference & $X$ & $Y$ & $Z$ & $B$ & Communication Time \\
\hline
\cite{ciciani1990analysis} & $M$ & $M$ & 1 & $+\infty$ & the average number of I/O’s per transaction $\times $ the average time per I/O \\
\hline
\cite{hwang1996data} & $M$ & $G$ & 1 & $+\infty$ & no \\
\hline
\cite{baccelli1982data} & $M$ & $M$ & m & $+\infty$ & no \\
\hline
\cite{bukh1992art} & $M$ & $D$ & $+\infty$ & $+\infty$ & constant \\
\hline
\cite{leung1997update} & $M$ & $H_{2}$ & 1 & $+\infty$ & constant 10 ms \\
\hline
our work & $M$ & $G$ & 1 & k & Rayleigh fading channel \\
\hline
\end{tabular}
\end{center}
\end{table*}

\section{SYSTEM}
\begin{figure}[h]\label{fig_1}
\centering
\includegraphics[width=3.5in]{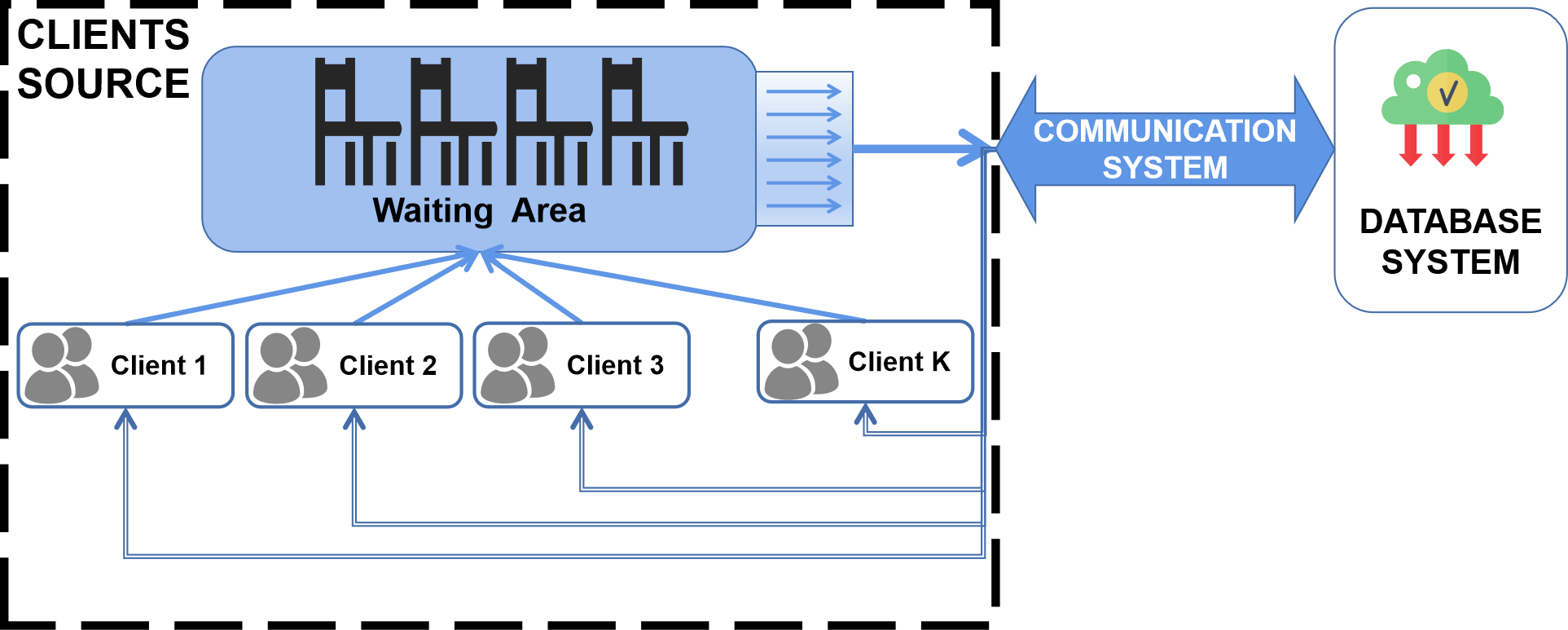}
\caption{C-D Model. The CLIENTS SOURCE, COMMUNICATION SYSTEM and DATABASE SYSTEM make up this model. The CLIENTS SOURCE is finite with $K$ clients, and has a Waiting Area. Quasi-static Rayleigh fading channel is considered in COMMUNICATION SYSTEM. And there is a single-server in the DATABASE SYSTEM.}
\end{figure}
\subsection{CLIENTS SOURCE}
This part assumes that there are $K(2<K<+\infty)$ clients. The client is locked until the completion of its final transaction. When a client is free at time $t$ (i.e., is not being served and is not waiting for service), it may initiate a transaction during interval $(t,t + \mathrm{d}t)$ with probability $\lambda$ dt.

If the channel is occupied or the DATABASE SYSTEM is locked, the transaction waits in the Waiting Area and generates a Poisson flow of repeated calls at a rate of $\gamma$ until it discovers that the channel is available. When the transaction waits in the Waiting Area, the client remains locked until terminated due to communication quality or successfully served.
\subsection{COMMUNICATION SYSTEM}
The channel in C-D Model is a quasi-static Rayleigh fading channel, such as \cite{5703199} and \cite{1198574}. It is assumed that fading coefficients are constant during a packet transmission, but change independently from one packet to another. If the transaction is terminated by database because of poor communication quality(i.e., the communication time exceeds $T$), it could be discarded. No data related to this discarded transaction is recorded.

We examine two cases in our work. In one case, $C=B \log _{2}\left(1+\frac{S}{N}\right)$, there is a nonlinear relationship between capacity and signal power in some SISO, called the nonlinear case. In another case, $C\approx \frac{S}{N\ln_{}{2} }$(see \cite{loyka2001channel}), there is a linear relationship between capacity and signal power in some MIMO, referred to as linear case.
\subsection{DATABASE SYSTEM}
Assume in the C-D Model that every transaction has an exclusive lock on this single-server database. Once a transaction enters the COMMUNICATION SYSTEM, the DATABASE SYSTEM is locked until this transaction finishes.

Poor communication quality leads to long communication times for transactions. It is presumed that $T$ represents the maximum time the database can wait for this transaction to transmit before terminating it. After the transaction is terminated, the database is unlocked for the next transaction.

The service time in the DATABASE SYSTEM conforms to an exponential distribution with a negative parameter $\mu$.
\subsection{The process of a transaction}
\begin{figure}[h]
\centering
\includegraphics[width=3.5in]{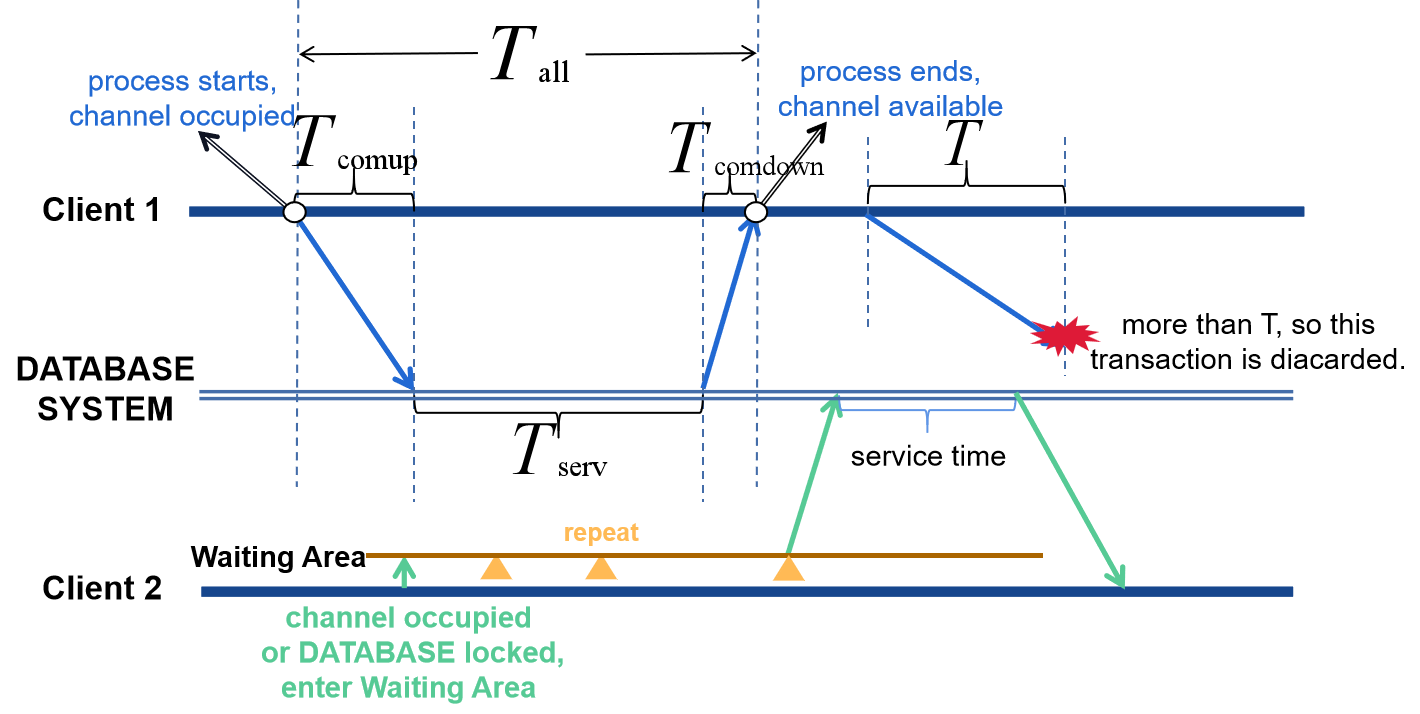}
\caption{The process of a transaction in CLIENTS SOURCE, COMMUNICATION SYSTEM and DATABASE SYSTEM.}
\label{fig_2}
\end{figure}
A transaction is initiated by one idle client in the CLIENTS SOURCE, which marks the beginning of a process. If the communication quality is so poor that the communication time is longer than $T$, the database system will terminate the transaction. This transaction is discarded and no data is recorded. On the contrary, if the DATABASE SYSTEM is unlocked in the case of good communication quality, transaction passes through the COMMUNICATION SYSTEM and is served by the DATABASE SYSTEM. Unfortunately, in the event that the channel is occupied or the DATABASE SYSTEM is locked, the transaction enters the Waiting Area and retries until the service successfully completes.

After the service is completed, the transaction passes through the COMMUNICATION SYSTEM again and returns to the CLIENTS SOURCE. In this case, the communication time longer than $T$ will also cause the transaction to be discarded. The client does not release the lock on the database until the returned transaction is received, which marks the end of the process.

The M/G/1//K queue is selected as the analytical model of our system based on the information presented above.
\section{ANALYSIS}
In this section, the M/G/1//K model is explained first. Secondly, the nonlinear and linear cases of ·/G/· are theoretically analyzed. These two cases have distinct distribution functions of $T_{all}$, where $T_{all} = T_{comup}+T_{serv}+T_{comdown}$. $T_{comup}$ is the uplink time in COMMUNICATION SYSTEM. $T_{serv}$ is the service time in DATABASE SYSTEM. $T_{comdown}$ is the downlink time in COMMUNICATION SYSTEM.
\subsection{M/G/1//K Model}
The following is an illustration of the model we used. The M/G/1//K queuing system was introduced by G. I. Falin and J. R. Artalejo in \cite{artalejo1998retrial}, and they improved the expressions of the main performance characteristics as proved in Appendix A$\footnote{The full paper is recorded on http://arxiv.org/abs/2212.09219.}$. Their results can be summarized as follows:
\begin{equation}\label{equ2}
q_{0 m}=C_{m} \cdot q_{0, K-1}, 0 \leq m \leq K-2,
\end{equation}
\begin{equation}\label{equ3}
q_{0, K-1}=v\left((v+\lambda+(K-1) \gamma) C_{0}+(\lambda-\gamma) C_{1}\right)^{-1},
\end{equation}
where $v=1 / \beta_{1}$, and
\begin{equation}\label{equ4}
\beta_{\mathrm{n}}=\int_{0}^{\infty} t^{n}  \mathrm{d} F_{T_{all}}(t).
\end{equation}

The coefficients $C_{m}$ can be recursively computed by eq(\ref{equ5}) with $q_{0, K-1}=1, q_{0 K}=q_{0,-1}=0$.
\begin{equation}\label{equ5}
\begin{aligned}
(((K-\mathrm{m}-1) \gamma+(\mathrm{m}+1) \lambda)(1-\beta(\mathrm{m} \lambda))+m \gamma \\
\beta(\mathrm{m} \lambda))q_{0 m}-(K-m) \gamma \beta(\mathrm{m} \lambda) q_{0, m-1}+(m+1)\\
(\lambda-\gamma)(1-\beta(\mathrm{m} \lambda)) q_{0, m+1}=0,1 \leq m \leq K-1,
\end{aligned}
\end{equation}
where $\beta \left ( s \right )$ is $F_{T_{all}}\left ( t \right )$'s Laplace-Stieltjes transform
\begin{equation}\label{equ6}
\beta(\mathrm{s})=\int_{0}^{\infty}e^{-s t}
  \mathrm{d} F_{T_{all}}(t).
\end{equation}
By using eq(\ref{equ2}), eq(\ref{equ5}) could be reduced to a equation with $C_{m}$. $C_{0}$ and $C_{1}$ are obtained by iteration. With eq(\ref{equ3}) and eq(\ref{equ2}), $q_{0, 0}$ is obtained.
When the system is in equilibrium,
\begin{enumerate}
\item[1)]The server utilization.
\begin{equation}\label{equ7}
\mathrm{p}_{1}=1-q_{00}.
\end{equation}
\item[2)]The mean number of clients in waiting area.
\begin{equation}\label{equ8}
 \mathrm{L}=K-\lambda^{-1}(\lambda+v) p_{1}.
\end{equation}
\item[3)]The mean waiting time.
\begin{equation}\label{equ9}
 \mathrm{W}=\left(v \mathrm{p}_{1}\right)^{-1} K-\lambda^{-1}-v^{-1}.
\end{equation}
\item[4)]The mean busy period.
\end{enumerate}

Busy periods are defined in this way, including alternating service periods and periods in which the DATABASE SYSTEM is unlocked and the Waiting Area has repeated transactions. The final results are given as eq(\ref{equ10}), and they are described in detail in \cite{falin1998finite} part4. 
\begin{equation}\label{equ10}
\begin{aligned}
\mathbf{E}\left [ L_{BP} \right ] =&\beta _{1} +  \left ( 1+\beta _{1}\left ( \left ( K- 1 \right )\mu +  \alpha\right)\right)\\
&\left (A_{0}\left ( 0 \right )+B_{0}\left ( 0 \right )\right )\\
&+ \left ( \alpha - \mu  \right ) \beta _{1}\left (A_{1}\left ( 0 \right ) +B_{1}\left ( 0 \right )\right ).
\end{aligned}
\end{equation}

$A_{m}\left ( 0 \right ),B_{m}\left ( 0 \right )$ are shown in eq(\ref{equ11}) and eq(\ref{equ12}).
\begin{figure*}
\begin{equation}\label{equ11}
\begin{split}
&A_{K-1} \left ( 0 \right ) = 0, B_{K-1} \left ( 0 \right ) = 0,\\
&A_{K-2} \left ( 0 \right ) = \left (  \mu \beta \left ( \left ( K- 1 \right )\alpha   \right )  \right ) ^{- 1},B_{K-2} \left ( 0 \right ) = - \mu ^{- 1},\\
&u_{m} \left ( 0 \right ) A_{m} \left ( 0 \right ) +v_{m} \left ( 0 \right ) A_{m+1} \left ( 0 \right )+ w_{m} \left ( 0 \right ) A_{m-1} \left ( 0 \right )=-\binom{K-1}{m},m=K-2,...,1,\\
&u_{m} \left ( 0 \right ) B_{m} \left ( 0 \right ) +v_{m} \left ( 0 \right ) B_{m+1} \left ( 0 \right )+ w_{m} \left ( 0 \right ) B_{m-1} \left ( 0 \right )=\binom{K-1}{m}\beta \left ( m\alpha  \right ) , m=K-2,...,1,
\end{split}
\end{equation}
\end{figure*}
\begin{figure*}
\begin{equation}\label{equ12}
\begin{split}
&u_{m}\left ( s \right ) = s+\left ( \left ( K-m-1 \right ) \mu +\left ( m+1 \right ) \alpha  \right )\left ( 1-\beta \left ( s+m\alpha  \right )  \right )+m\mu \beta \left ( s+m\alpha  \right ) ,  m=K-1,...,1,\\
&v_{m}\left ( s \right ) = \left ( m+1 \right ) \left ( \alpha -\mu  \right ) \left ( 1-\beta \left ( s+m\alpha  \right )  \right ), m=K-1,...,1,\\
&w_{m}\left ( s \right ) = \left ( K-m \right ) \mu \beta \left ( s+m\alpha  \right ) ,  m=K-1,...,1.
\end{split}
\end{equation}
\end{figure*}
\subsection{Distribution Functions of $T_{all}$}
The Rayleigh fading model is suitable for describing the wireless channel and belongs to small-size fading. In this paper, the large-scale fading effect does not affect the theoretical derivation, so that it can be ignored. The probability density function of the Rayleigh distribution is shown in eq(\ref{equ13}).
\begin{equation}\label{equ13}
f\left ( r \right ) =2\alpha re^{-\alpha r^{2}}, r\in(0,+\infty),
\end{equation}
where $\alpha=\frac{1}{2\sigma ^{2}}$, $\sigma$ is the parameter in Rayleigh fading channel .

According to eq(\ref{equ13}), the distribution function of the signal power is deduced in eq(\ref{equ14}).
\begin{equation}\label{equ14}
\begin{aligned}
F_{S} \left ( s \right ) &=P\left \{ R\le \sqrt{s} \right \}\\
&=1-e^{-\alpha s}, s \in(0,+\infty).
\end{aligned}
\end{equation}

It is assumed that the amount of information that goes through the channel each time is a unit packet, which is 1KB\cite{10.1145/1410043.1410045}. 
\subsubsection{Linear Case}
In the linear case, there is a linear relationship between capacity and signal power in some MIMO. Under the information parallel transmission technology in \cite{loyka2001channel}, the capacity meets $C\approx \frac{S}{N\ln_{}{2} }$. Therefore, $T_{com}$ is deduced as follows.
\begin{equation}\label{equ15}
T_{com}=\frac{1}{C}=\frac{N\ln_{}{2} }{S} .
\end{equation}

According to eq(\ref{equ14}) and eq(\ref{equ15}), the probability density function $f_{T_{com}}(t)$ of the communication time $T_{comup}$ and $T_{comdown}$ is shown in eq(\ref{equ16}).
\begin{equation}\label{equ16}
\begin{aligned}
f_{T_{com} } \left ( t \right ) = \frac{N\ln_{}{2}\cdot \alpha}{t^{2} } e^{-\frac{N\ln_{}{2}\cdot \alpha }{t} },t \in(0,T].
\end{aligned}
\end{equation}

The service time follows exponential distribution with a negative parameter $\mu$. So the probability density function $f_{T_{serv}}(t)$ of the service times $T_{serv}$ is
\begin{equation}\label{equ17}
f_{T_{serv}}(t)=\mu e^{-\mu t}, t \in(0,+\infty).
\end{equation}

First, it is assumed that $Y=T_{comup}+T_{serv}$. As proved in Appendix B$\footnote{The full paper is recorded on http://arxiv.org/abs/2212.09219.}$, we arrive at
\begin{equation}\label{equ18}
f_{Y}(y)=\left\{\begin{array}{l}
N\ln_{}{2}\alpha \mu e^{-\mu y}g\left ( y \right ), y \in(0,T],
\\ N\ln_{}{2}\alpha \mu e^{-\mu y}g\left ( T \right ), y \in(T,\infty ),
\\0, else,
\end{array}\right.
\end{equation}
where
\begin{equation}\label{equ19}
\begin{aligned}
&g\left ( y \right )=\int_{\frac{1}{y}}^{\infty}e^{-N\ln_{}{2}\cdot \alpha r +\frac{\mu }{r}}  \mathrm{d}r=\\
&\frac{ S\left [ \frac{-1}{N\ln_{}{2}\cdot \alpha \mu } ,\frac{y}{N\ln_{}{2}\cdot  \alpha }  \bigg|\begin{bmatrix}
1,0\\0,0\end{bmatrix} _{-}^{1} \bigg| \begin{pmatrix}0,1\\1,0\end{pmatrix}\begin{matrix}
1\\-\end{matrix} \bigg| \begin{pmatrix}0,1\\1,1\end{pmatrix}\begin{matrix}
1\\0\end{matrix}\right ]}{N\ln_{}{2}\cdot \alpha  }.
\end{aligned}
\end{equation}

\begin{figure*}
\begin{equation}\label{equ20}
F_{T_{all}}(t)=\left\{\begin{array}{l}
e^{-\frac{2N\ln_{}{2}\cdot \alpha}{t}}
H\begin{matrix}0,2\\2,0\end{matrix}\left [ \frac{t^{2}}{\left (N\ln_{}{2}\cdot \alpha \right )^{2}} \bigg| \begin{matrix}
(1,1),(0,1)\\-\end{matrix} \right ]-N\ln_{}{2}\cdot \alpha e^{-\mu t}\int_{0}^{+\infty} e^{-N\ln_{}{2}\cdot \alpha x}
\\S\left [ \frac{x}{\mu } ,tx \bigg|\begin{bmatrix}
0,0\\0,0\end{bmatrix}_{-}^{-}\begin{pmatrix}0,1\\1,0\end{pmatrix}\begin{matrix}
1\\-\end{matrix}\begin{pmatrix}0,1\\1,1\end{pmatrix}\begin{matrix}
1\\0\end{matrix}\right ]
S\left [- \frac{1}{N\ln_{}{2}\cdot \alpha\mu } ,\frac{1}{N\ln_{}{2}\cdot \alpha \left ( t-x \right ) } \bigg| \begin{bmatrix}
1,0\\0,0\end{bmatrix}_{1}^{-} \bigg|\begin{pmatrix}0,1\\1,0\end{pmatrix}\begin{matrix}
1\\-\end{matrix} \bigg|\begin{pmatrix}0,1\\1,1\end{pmatrix}\begin{matrix}
1\\0\end{matrix}\right ]\mathrm{d}x,
\\\qquad \qquad \qquad \qquad \qquad \qquad \qquad \qquad \qquad \qquad \qquad \qquad \qquad \qquad \qquad \qquad \qquad \qquad t \in(0,T],
\\e^{ -\frac{2N\ln_{}{2}\cdot \alpha }{t}}\left \{
S\left [ \frac{t^{2}}{\left (N\ln_{}{2}\cdot \alpha \right )^{2}},\frac{t\cdot T}{N\ln_{}{2}\cdot \alpha \left ( t-T \right ) }  \bigg| \begin{bmatrix}
1,0\\0,0\end{bmatrix} _{-}^{1} \bigg| \begin{pmatrix}0,1\\1,0\end{pmatrix}\begin{matrix}
1\\-\end{matrix} \bigg| \begin{pmatrix}0,1\\1,1\end{pmatrix}\begin{matrix}
1\\0\end{matrix} \right ]
\right.
\\
\phantom{=\;\;}
\left.-S\left [\frac{t^{2}}{\left (N\ln_{}{2}\cdot \alpha \right )^{2}} ,\frac{t\left ( t-T \right )}{N\ln_{}{2}\cdot \alpha T}  \bigg| \begin{bmatrix}
1,0\\0,0\end{bmatrix} _{-}^{1} \bigg| \begin{pmatrix}0,1\\1,0\end{pmatrix}\begin{matrix}
1\\-\end{matrix} \bigg| \begin{pmatrix}0,1\\1,1\end{pmatrix}\begin{matrix}
1\\0\end{matrix}\right ] \right \}
-N\ln_{}{2}\cdot \alpha e^{-\mu t}\int_{0}^{+\infty} e^{N\ln_{}{2}\cdot \alpha x}
\\S\left [ \frac{x}{\mu } ,Tx \bigg| \begin{bmatrix}
0,0\\0,0\end{bmatrix} _{-}^{-} \bigg| \begin{pmatrix}0,1\\1,0\end{pmatrix}\begin{matrix}
1\\-\end{matrix} \bigg| \begin{pmatrix}0,1\\1,1\end{pmatrix}\begin{matrix}
1\\0\end{matrix}\right ]
\\S\left [- \frac{1}{N\ln_{}{2}\cdot \alpha \mu } ,\frac{1}{N\ln_{}{2}\cdot \alpha \left ( t-x \right ) }  \bigg| \begin{bmatrix}
1,0\\0,0\end{bmatrix} _{-}^{1} \bigg| \begin{pmatrix}0,1\\1,0\end{pmatrix}\begin{matrix}
1\\-\end{matrix} \bigg| \begin{pmatrix}0,1\\1,1\end{pmatrix}\begin{matrix}
1\\0\end{matrix}\right ] \mathrm{d}x
+N\ln_{}{2}\cdot \alpha e^{-\mu t}\int_{0}^{+\infty} e^{-N\ln_{}{2}\cdot \alpha x}
\\S\left [ \frac{x}{\mu } ,(t-T )x \bigg| \begin{bmatrix}
0,0\\0,0\end{bmatrix} _{-}^{-} \bigg| \begin{pmatrix}0,1\\1,0\end{pmatrix}\begin{matrix}
1\\-\end{matrix} \bigg| \begin{pmatrix}0,1\\1,1\end{pmatrix}\begin{matrix}
1\\0\end{matrix}\right ]
\\S\left [- \frac{1}{N\ln_{}{2}\cdot \alpha \mu } ,\frac{1}{N\ln_{}{2}\cdot \alpha \left ( t-x \right ) }  \bigg| \begin{bmatrix}
1,0\\0,0\end{bmatrix} _{-}^{1} \bigg| \begin{pmatrix}0,1\\1,0\end{pmatrix}\begin{matrix}
1\\-\end{matrix} \bigg| \begin{pmatrix}0,1\\1,1\end{pmatrix}\begin{matrix}
1\\0\end{matrix}\right ]  \mathrm{d}x
\\+N\ln_{}{2}\cdot \alpha e^{ -\mu T-\frac{N\ln_{}{2}\cdot \alpha}{t-T}}g\left ( T \right )-\left ( N\ln_{}{2}\cdot \alpha \right ) ^{2}e^{-\mu t }g\left (T\right )g\left ( t-T\right)+ h\left (T \right )
e^{-\frac{\alpha}{t-T}}, t \in(T,2T],
\\e^{- \frac{2N\ln_{}{2}\cdot \alpha }{T }} -\left ( N\ln_{}{2}\cdot \alpha \right )^{2} e^{ -\mu t } g\left ( T \right ), t\in(2T,+\infty),\\
0, else,
\end{array}\right.
\end{equation}
\end{figure*}
Second, $Z(i.e., T_{all})$ is defined as $Z=T_{comup}+T_{serv}+T_{comdown}=Y+T_{comdown}$, $y\in(0,+\infty), t_{comdown}\in(0,T]$. The calculation is the same as for $Y$. Then, in linear case, the distribution function $F_{T_{all}} (t)$ is obtained in eq(\ref{equ20}), where
\begin{equation}\label{equ21}
\begin{aligned}
h\left ( T \right )&=e^{ -\frac{\alpha }{T}}-\alpha e^{  -\mu T} g\left (T \right ) .
\end{aligned}
\end{equation}

$F_{T_{all}} (t)$ is divided into three segments according to its domain. When $t$ is less than $2T$, $F_{T_{all}} (t)$ grows rapidly, that is, the total service time of a customer is around $2T$. $F_{T_{all}} (t)$ eventually approaches maximum.

\subsubsection{Nonlinear Case}
In nonlinear case, there is a nonlinear relationship between capacity and signal power in some SISO.

First of all, the probability density function of the service times $T_{serv}$, in this case, is as same as in eq(\ref{equ17}) because the DATABASE SYSTEM is as same as the nonlinear case. According to capacity $C=B \log _{2}\left(1+\frac{S}{N}\right)$, we arrive at
\begin{equation}\label{equ22}
T_{com}=\frac{1}{C}=\frac{1}{B\log_{2}{\left ( 1+\frac{S}{N}  \right ) } }.
\end{equation}

In Rayleigh fading channel, $T_{comup}$ and $T_{comdown}$ have the same distribution function $F_{T_{com}}(t)$ which is deduced from eq(\ref{equ14}) and eq(\ref{equ22}).
\begin{equation}\label{equ23}
\begin{aligned}
F_{T_{com} } \left ( t \right ) &=P\left \{ S\ge N\left ( 2^{\frac{1}{Bt} } -1 \right )  \right \} \\
&=e^{-\alpha N\left ( 2^{\frac{1}{Bt} } -1 \right )}, t \in(0,T].
\end{aligned}
\end{equation}

According to eq(\ref{equ17}) and eq(\ref{equ23}), the distribution function $F_{T_{all}} (t)$ of time $T_{all}=T_{comup}+T_{serv}+T_{comdown}$ is shown in eq(\ref{equ24}).
\begin{equation}\label{equ24}
F_{T_{all}} (t)=\left\{\begin{array}{l}
\int_{0}^{z} f_{T_{com} } \left ( x \right )F_{1} \left ( t-x \right ) \mathrm{d}x ,t \in(0,T],
\\\int_{0}^{t-T} f_{T_{com} } \left ( x \right )F_{2} \left ( t-x \right ) \mathrm{d}x\\
+\int_{t-T}^{T} f_{T_{com} } \left ( x \right )F_{1} \left ( t-x \right ) \mathrm{d}x,\\
\qquad \qquad \qquad \qquad \qquad t \in(T,2T],
\\\int_{0}^{T} f_{T_{com} } \left ( x \right )F_{2} \left ( t-x \right ) \mathrm{d}x, \\
\qquad \qquad \qquad \qquad t \in(2T,+\infty),\\
0, else.
\end{array}\right.
\end{equation}

\emph{Notation:} In eq(\ref{equ24}), it is assumed that $F_{1} \left ( x \right )=F(x),x \in(0,\mathrm{T}]  $ and $F_{2} \left ( x \right )=F(x),x \in(\mathrm{T},+\infty) $. The $F(x)(i.e., X=T_{comup}+T_{serv})$ is shown in eq(\ref{equ25}).
\begin{equation}\label{equ25}
F(x)=\left\{\begin{array}{l}
\mu e^{-\mu x+\alpha N} \int_{0}^{x} e^{-\mu t-\alpha N \cdot 2^{\frac{1}{Bt}}} \mathrm{d}t,
\\\qquad \qquad \qquad \qquad x \in(0,T],
\\F\left ( T \right )\left ( 1-e^{\mu T- \mu x} \right )+\mu  e^{-\mu x+\alpha N}
\\ \cdot \int_{0}^{T} e^{\mu t-\alpha N \cdot 2^{\frac{1}{B t}}}\mathrm{d}t , x \in(T,+\infty),
\\0, else.
\end{array}\right.
\end{equation}

\section{SIMULATION RESULTS}
This section describes the parameters of procedure code and results of the C-D Model in figures. The lines in our figures represent the theoretical results gained from the section \uppercase\expandafter{\romannumeral3}, while the marks represent the simulation results received from the queue program.

The results will be presented in three parts. The first is the distribution function $F_{T_{all}} (t)$ in two cases, which is the correctness verification of the mathematical derivation. The second group consists of client-related metrics including L and W. The others, WS and LS are shown in Appendix C$\footnote{The full paper is recorded on http://arxiv.org/abs/2212.09219.}$. The third aspect is the BP, where observe the effectiveness of the C-D Model from the standpoint of the database.
\subsection{PDF:The probability density function of $T_{all}$}
The values of the parameters are set as follows: In linear case, $\mu$=1/10, $T$=3, $\alpha$=3. In nonlinear case, $\mu$=1/10, $T$=3, $\alpha$=3, $B$=1, $N$=2.
\begin{figure}[ht]
\centering
\includegraphics[width=3.5in]{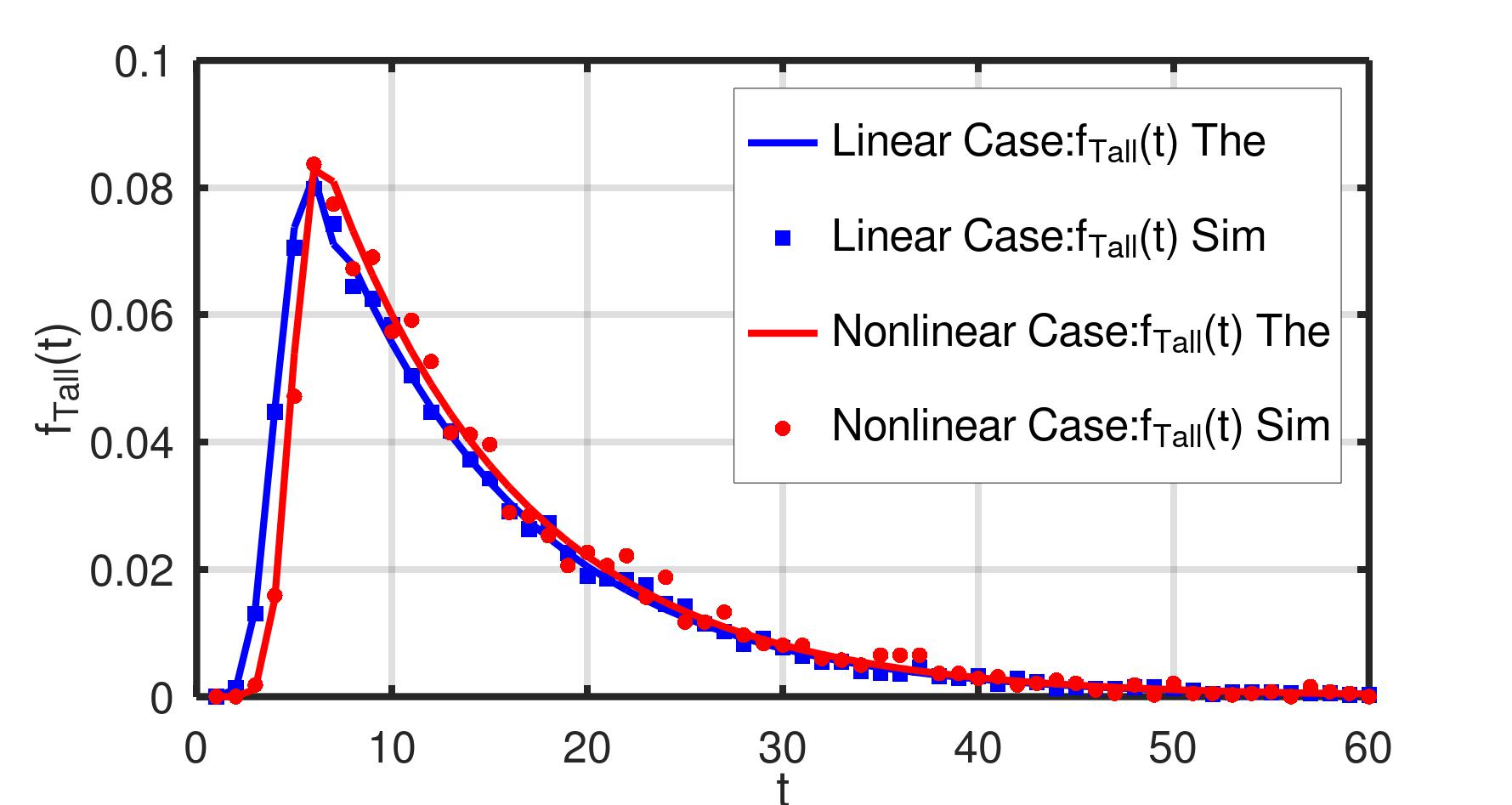}
\caption{Probability Density Function of time $T_{all}=T_{comup}+T_{serv}+T_{comdown}$.}
\label{fig_3}
\end{figure}

In section \uppercase\expandafter{\romannumeral3}, the probability density function of $T_{all}$ is calculated in both nonlinear and linear case. In Fig. \ref{fig_3}, the results we get from eq(\ref{equ20}) and eq(\ref{equ24}) are shown in lines, while marks represent the simulation results. Evidently, our analysis corresponds to the simulation. Moreover, the $t$ is concentrated around a smaller number in the linear case. This illustrates that the communication quality is indeed improved under some MIMO technologies.
\subsection{Different $\alpha$ of Rayleigh Fading Channel}
In this part, the values of the parameters are set as follows: $K$=10, $\mu$=1/10, $T$=3, nonlin-$B$=1, nonlin-$N$=1, $\gamma$=1/2.

\emph{Notation:} The abscissa, $1/\lambda$, represents the average time interval between client arrivals. When $1/\lambda$ increases, the density of client arrival decreases.

\begin{figure}[ht]
\centering
\includegraphics[width=3.5 in]{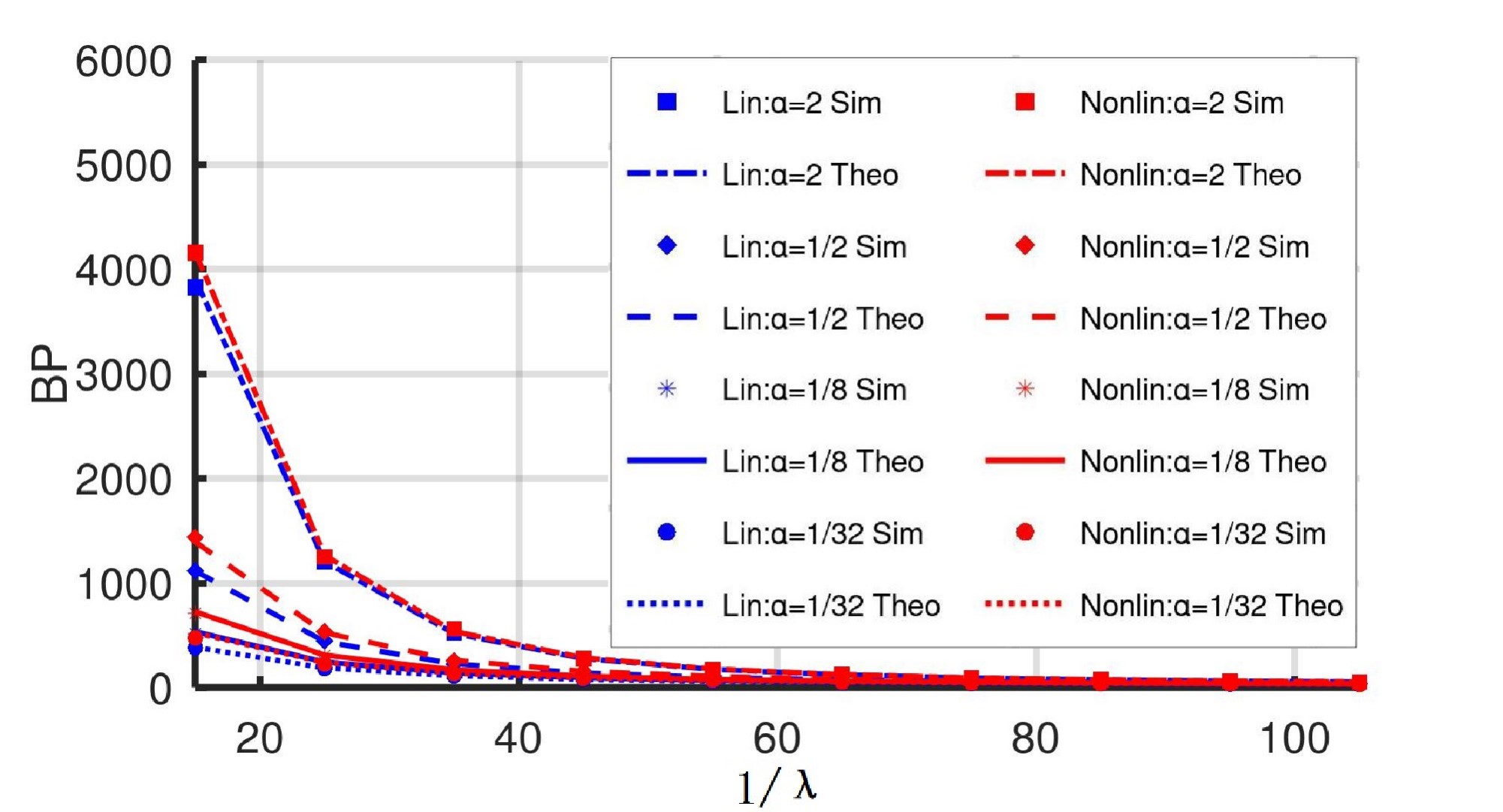}
\caption{The busy time(i.e., consists of alternating service periods and periods during which the Service Area is unlocked and there are repeated transactions in the Waiting Area) of DATABASE SYSTEM with different $\alpha$.}
\label{fig_4}
\end{figure}
Fig. \ref{fig_4} presents the results of the busy period of the DATABASE SYSTEM, which is the only metric from the database perspective. First of all, the theoretical results of the model are compatible with its simulation results. The observations show that better communication quality(i.e., smaller $\alpha$) facilitates the reduction of BP. These lines decrease fast as $1/\lambda$ increases in both cases. In addition, using the exponential correlation matrix(i.e., MIMO in \cite{loyka2001channel}) does ease the busyness of the DATABASE SYSTEM.

\begin{figure}[ht]
\centering
\includegraphics[width=3.5in]{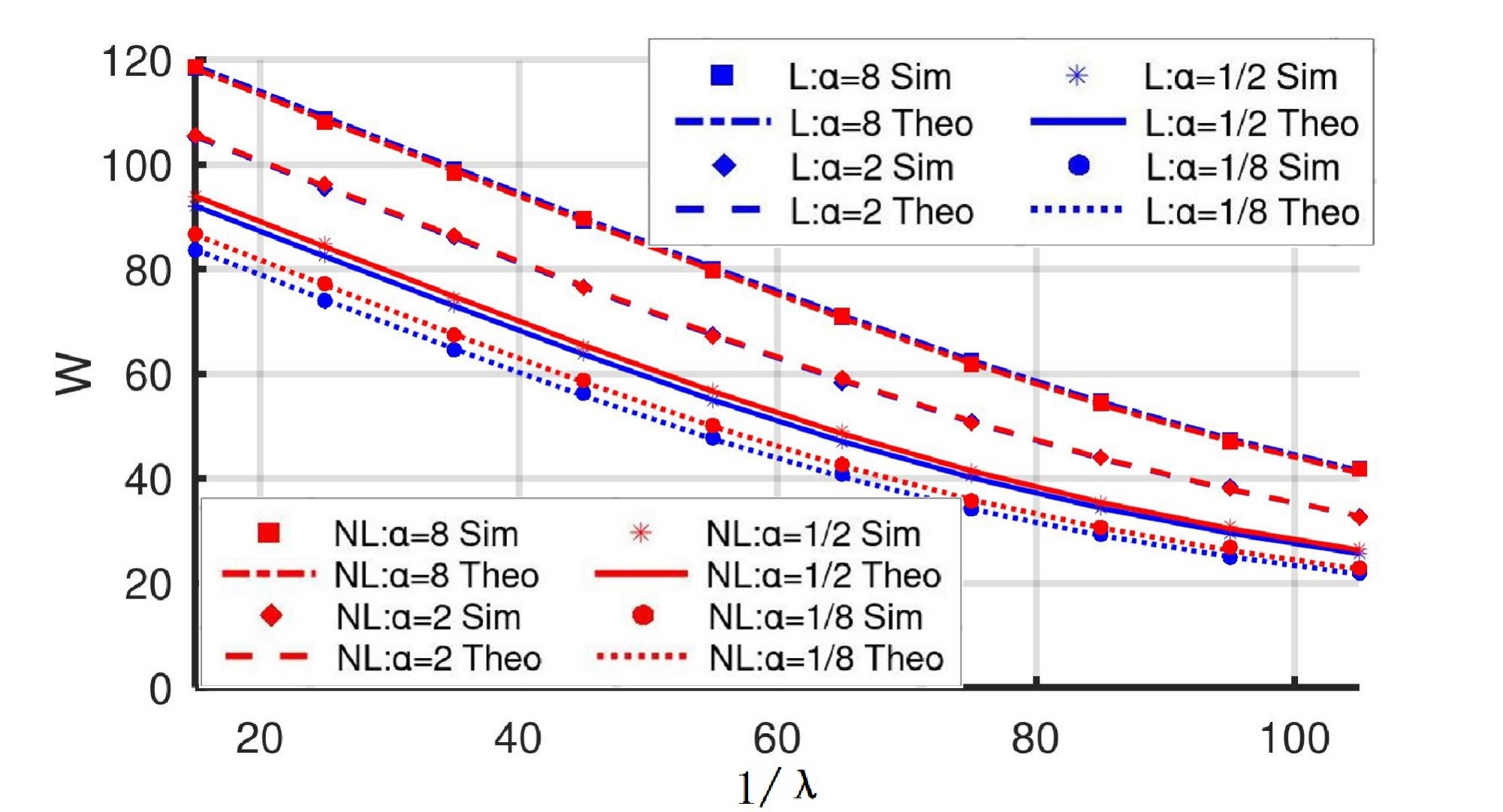}
\caption{Different $\alpha$: Theoretical(from equation) and simulation(from code) results on W.}
\label{fig_5}
\end{figure}
Fig. \ref{fig_5} presents the results of the W. In both cases, W increases as communication quality declines. And from a lateral perspective, the more intensive the transactions(i.e., smaller $1/\lambda$), the more clogged the system(i.e., bigger W). The most important one is that the theoretical results from equations are consistent with the simulation results, indicating means our model works.
\subsection{ Different B and N  of Rayleigh Fading Channel(Nonlinear Case)}
In this part, the values of the parameters are set as follows: $K$=10, $\mu$=1/10, $T$=3, $\alpha$=1/2, $\gamma$=1/2 and initial value: $B$=1, $N$=1.
\begin{figure}[ht]
\centering
\includegraphics[width=3.5in]{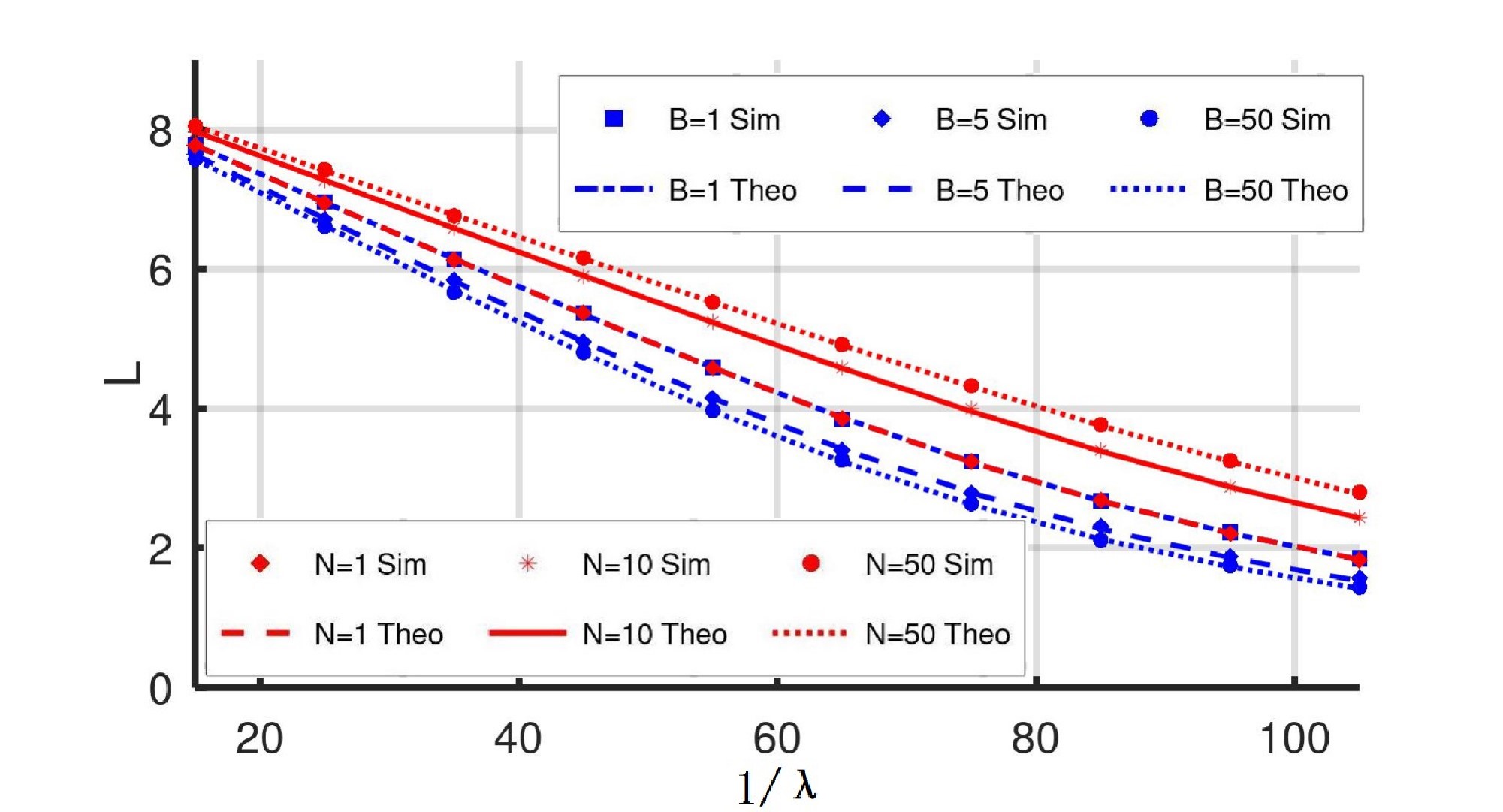}
\caption{Different $B$ and $N$: Theoretical(from equation) and simulation(from code) results on L.}
\label{fig_6}
\end{figure}

Fig. \ref{fig_6} depicts the metrics of L in different $B$ and $N$ settings. It can be seen that the theoretical results of the model and simulation results match each other. In addition to L decreasing as transaction density decreases, it can also be seen that the spacing between lines decreases as $B$ increases. And once the bandwidth increases to a certain extent, neither the lowering of L nor the optimization of the system are significantly affected by it. As $N$ reduces, L also reduce at the same $1/\lambda$. It simply demonstrates that channel quality with low noise can reduce database processing congestion. 
\section{Conclusion}
This paper proposes a new model, named the C-D Model, based on communication quality and the service capability of the database system, in which theory and simulation are complementary. The M/G/1//K is selected to describe the database system. The C-D Model can be used to study online services under various communication qualities with the assumption of the linear and nonlinear cases. Future work will consider optimizing the C-D Model by adopting more complex channel models and database processing models closer to real applications, such as M/G/c//K. Combining the C-D Model with cloud computing produces observation models for improving the performance of cloud computing.

%

\bibliographystyle{IEEEtran}
\bibliography{mylib}


\appendices

\section{The derivation of M/G/1//K}
In \cite{artalejo1998retrial}, the $C(t)$ is 0 or 1 according as the server is free or busy at time $t$, $N(t)$ is the number of sources in orbit. The length of the busy periods will be denoted by $L_{BP}$. The probabilities(densities) are defined as follows:
\begin{equation}\label{equ26}
\begin{split}
p_{0 n}=\mathbf{P}\{L_{BP}>t, C(t)=0, N(t)=n\}, \\
\quad 0 \leq n \leq K-1 ,
\end{split}
\end{equation}
\begin{equation}\label{equ27}
\begin{aligned}
p_{1 n}(x)=\frac{ \mathrm{d}}{ \mathrm{d} x} \mathbf{P}\{L_{BP}>t, C(t)=1, \xi(t) \leq x, \\
N(t)=n\}, \quad 0 \leq n \leq K-1 ,
\end{aligned}
\end{equation}
\begin{equation}\label{equ28}
\begin{split}
p_{1 n}&=\mathbf{P}\{L_{BP}>t, C(t)=1, N(t)=n\}\\
&=\int_{0}^{\infty} p_{1 n}(x)  \mathrm{d} x, \quad 0 \leq n \leq K-1,
\end{split}
\end{equation}
where $\xi(t)$ is the service time that has lasted when $C(t)=1$.

Then, following the method of supplementary variables, the limiting probabilities as $t$ approaches $+\infty$  satisfy the equations of statistical equilibrium:
\begin{equation}\label{equ29}
((K-n) \alpha+n \mu) p_{0n}=\int_{0}^{\infty} p_{1n}(x) f_{T_{all}}(x)  \mathrm{d} x ,
\end{equation}
\begin{equation}\label{equ30}
\begin{aligned}
p_{1 n}^{\prime}(x)=&-((K-n-1) \alpha+f_{T_{all}}(x)) p_{1 n}(x)\\
&+(K-n) \alpha p_{1, n-1}(x) ,
\end{aligned}
\end{equation}
where $f_{T_{all}}(x)=F_{T_{all}}^{\prime}(x) /(1-F_{T_{all}}(x))$ is the hazard rate function of $F_{T_{all}}(x)$ and $p_{0K}=p_{1,-1}=0$. With the help of so-called discrete transformations in \cite{artalejo1998retrial}, a set of unknown variables $p=(p_{0},...,p_{K-1})$ is replaced by $q^{\prime}=(q_{0},...,q_{K-1})^{\prime}= Ap^{\prime}$, where $A$ is a non-singular $K\times K$ matrix.
\begin{equation}\label{equ31}
\begin{split}
p_{n}=\sum_{m=0}^{n}(-1)^{m}\left(\begin{array}{c}
K-1-n+m \\
m
\end{array}\right) q_{K-1-n+m},\\
 \quad 0 \leq n \leq K-1.
\end{split}
\end{equation}

\section{The derivation of the probability density function for Y}
It is assumed that $Y=T_{comup}+T_{serv}$, (i.e.In order to make it easier to write, it is assumed that $T_{comup}=T_{comdown}=T_{com}=U, T_{serv}=X $), $u(i.e.t_{comup})\in(0,T],x(i.e.t_{serv})\in(0,+\infty)$.

When $Y\le 0$, $F_{Y}\left ( y \right )=0$. When $0< Y\le T$,
\begin{equation}\label{equ32}
\begin{aligned}
F_{Y}\left ( y \right ) =e^{-\frac{N\ln_{}{2}\cdot \alpha }{y} }-e^{-\mu y}\\
\int_{0}^{y}\frac{N\ln_{}{2}\cdot \alpha e^{-\frac{N\ln_{}{2}\cdot \alpha }{u} +\mu u }}{u ^{2} }  \mathrm{d}u, y \in(0,T].
\end{aligned}
\end{equation}
\begin{equation}\label{equ33}
\begin{aligned}
f_{Y}\left ( y \right ) =N\ln_{}{2}\alpha \mu e^{-\mu y}\int_{\frac{1}{y}}^{\infty}e^{-N\ln_{}{2}\cdot \alpha r +\frac{\mu }{r}}  \mathrm{d}r, \\
y \in(0,T].
\end{aligned}
\end{equation}

When $Y>T$,
\begin{equation}\label{equ34}
\begin{aligned}
F_{Y}\left ( y \right )=&e^{-\frac{N\ln_{}{2}\cdot \alpha }{T} }\\
&-e^{-\mu y}\int_{0}^{T}\frac{N\ln_{}{2}\cdot \alpha e^{-\frac{N\ln_{}{2}\cdot \alpha }{u} +\mu u }}{u ^{2} }  \mathrm{d}u, \\
&\qquad \qquad \qquad \qquad \qquad y \in(T,\infty ).
\end{aligned}
\end{equation}
\begin{equation}\label{equ35}
\begin{aligned}
f_{Y}\left ( y \right ) =N\ln_{}{2}\alpha \mu e^{-\mu y}\int_{\frac{1}{T}}^{\infty}e^{-N\ln_{}{2}\cdot \alpha r +\frac{\mu }{r}}  \mathrm{d}r, \\
 y \in(T,\infty ).
\end{aligned}
\end{equation}

\section{figures of theoretical and simulation results}
\begin{figure}[ht]
\centering
\includegraphics[width=3.5in]{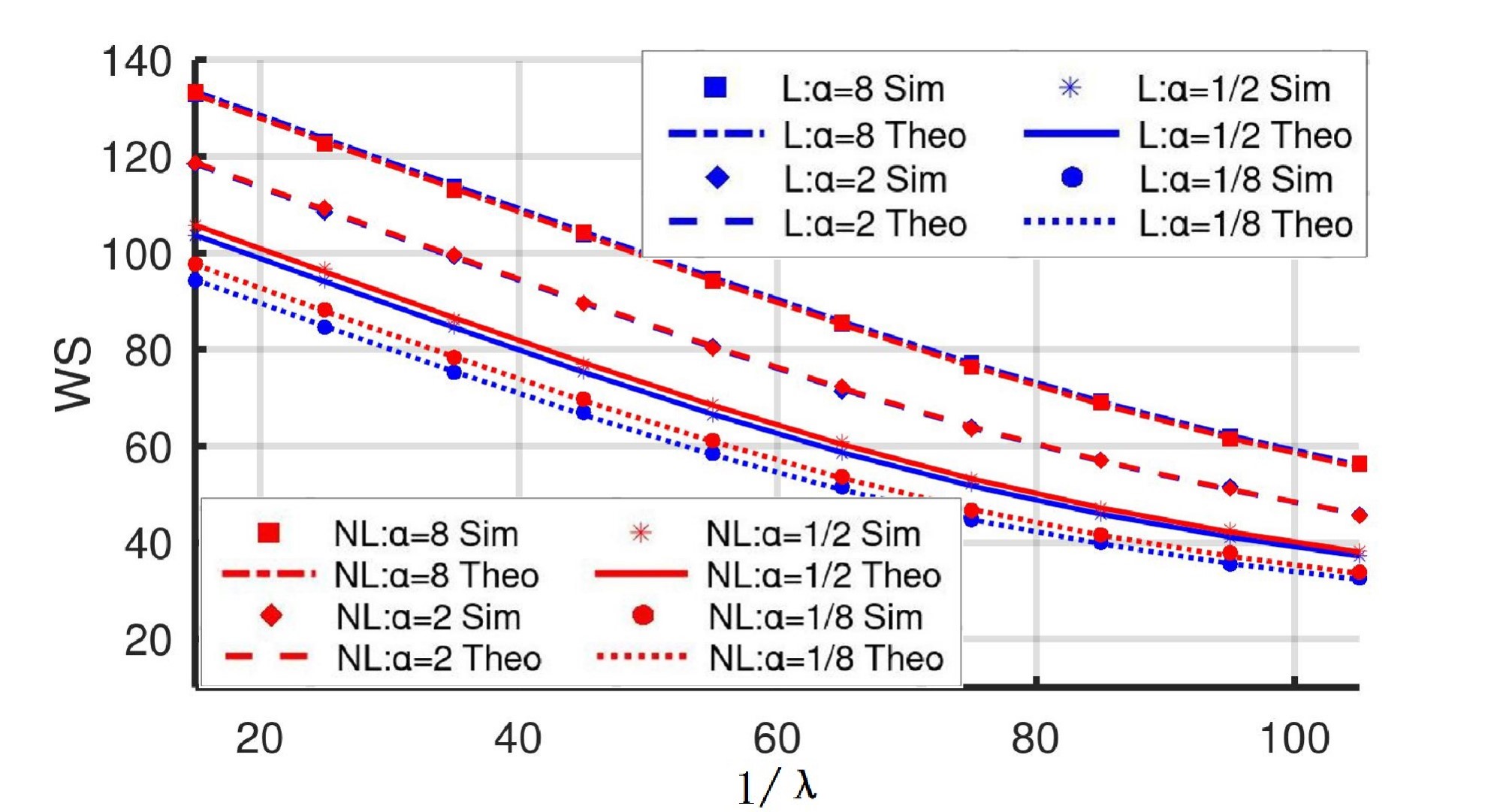}
\caption{Different $\alpha$: Theoretical(from equation) and simulation(from code) results on WS.}
\label{fig_7}
\end{figure}
\begin{figure}[ht]
\centering
\includegraphics[width=3.5in]{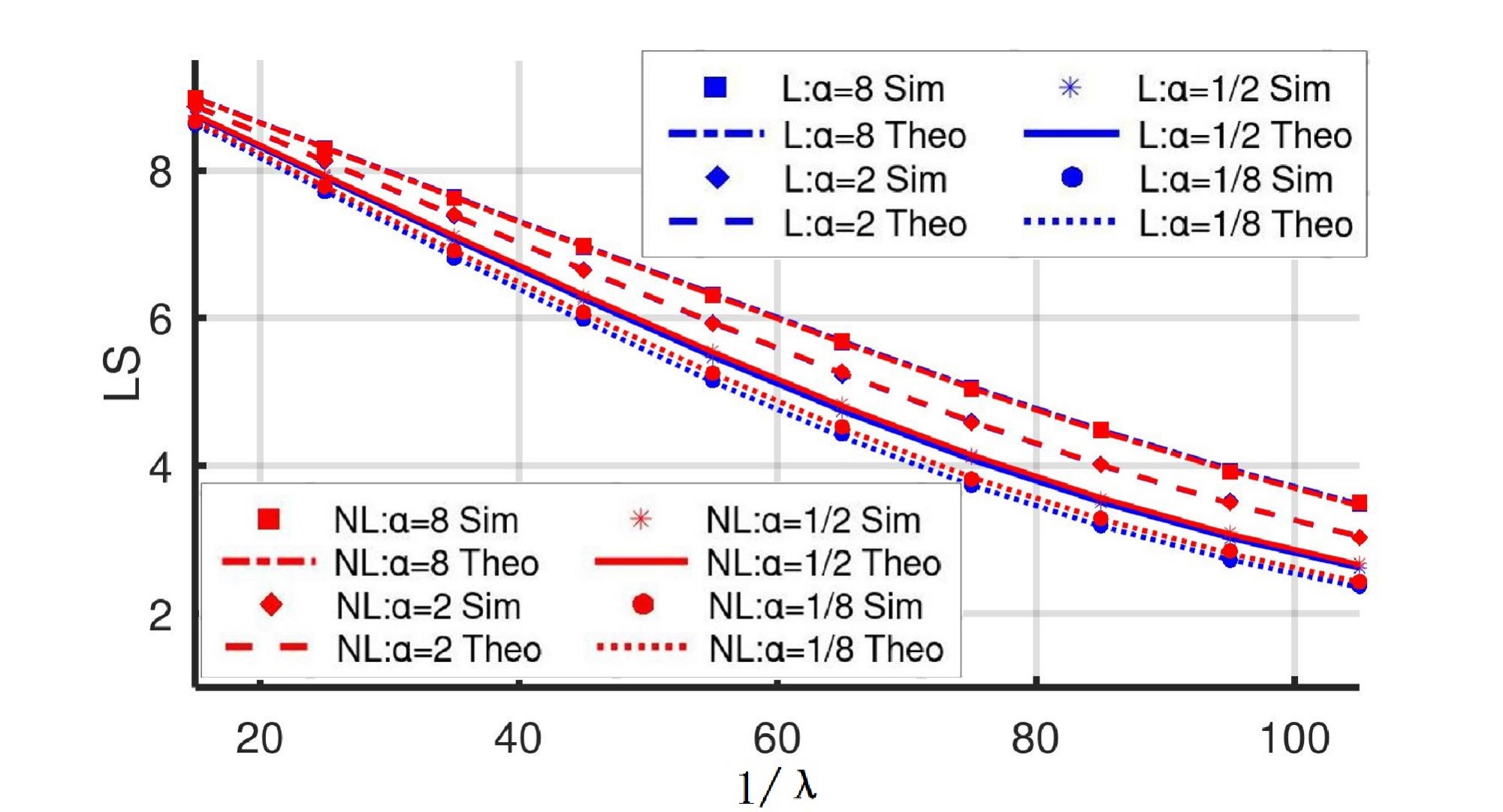}
\caption{Different $\alpha$: Theoretical(from equation) and simulation(from code) results on LS.}
\label{fig_8}
\end{figure}
\begin{figure}[ht]
\centering
\includegraphics[width=3.5in]{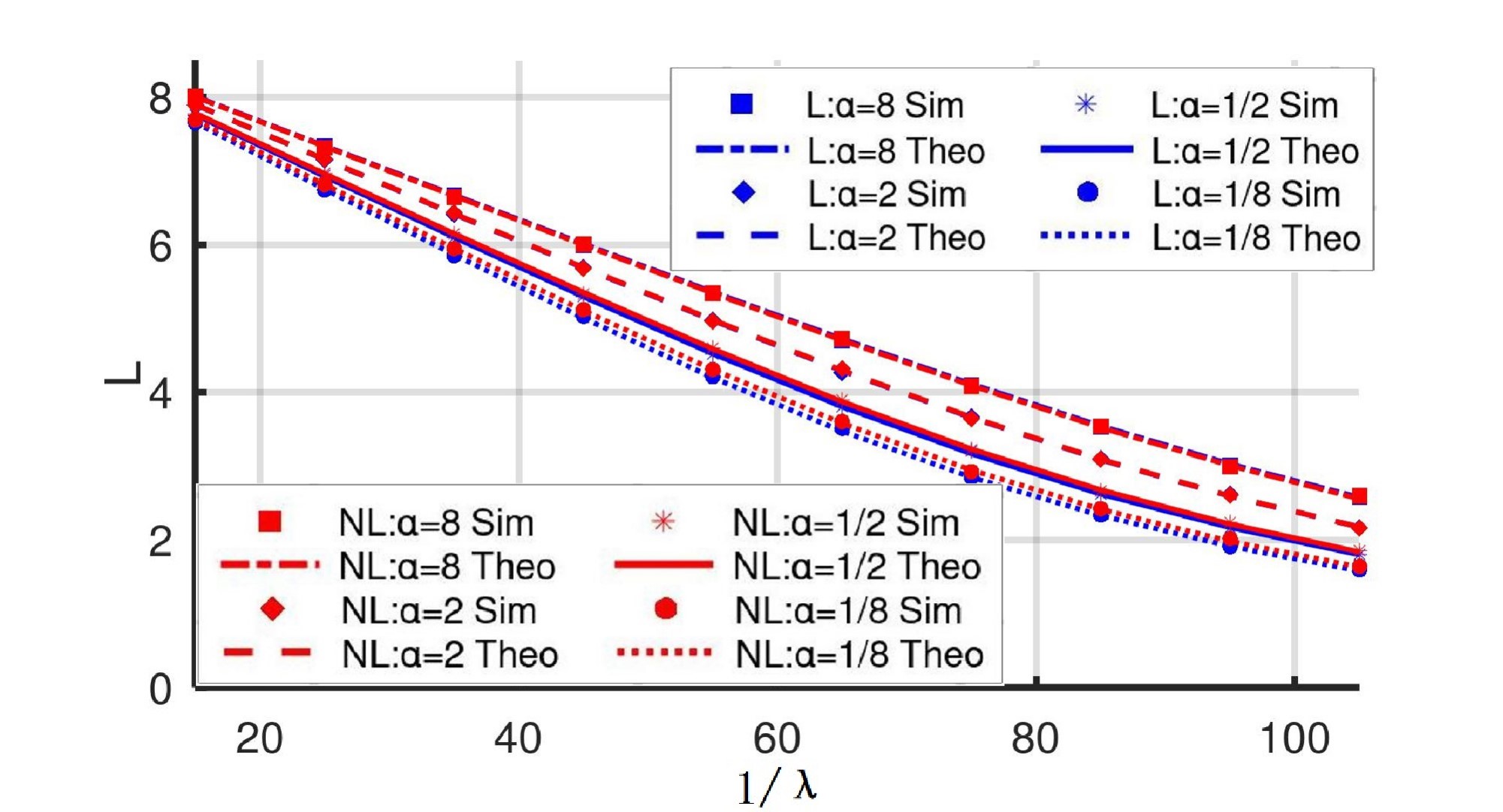}
\caption{Different $\alpha$: Theoretical(from equation) and simulation(from code) results on L.}
\label{fig_9}
\end{figure}

In Fig. \ref{fig_7}, Fig. \ref{fig_8} and Fig. \ref{fig_9}, all lines increase as communication quality declines. And from a lateral perspective, the more intensive the transactions(i.e., smaller $1/\lambda$), the more clogged the system(i.e., bigger WS, LS and L). The most important one is that the analytic results from equations are consistent with the simulation results, indicating means our model works.
\begin{figure}[ht]
\centering
\includegraphics[width=3.5in]{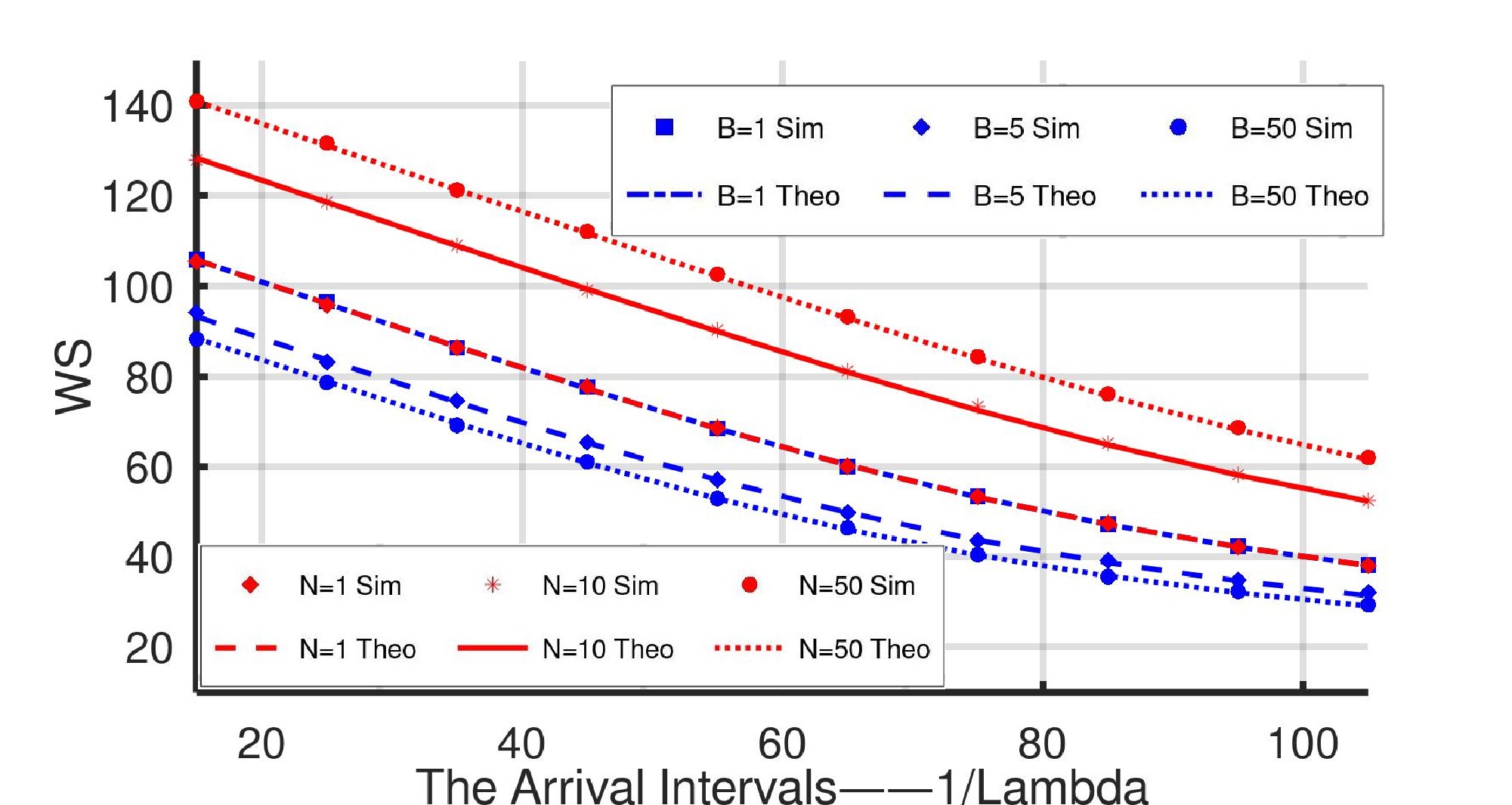}
\caption{Different B and N: Theoretical(from equation) and simulation(from code) results on WS.}
\label{fig_10}
\end{figure}

\begin{figure}[ht]
\centering
\includegraphics[width=3.5in]{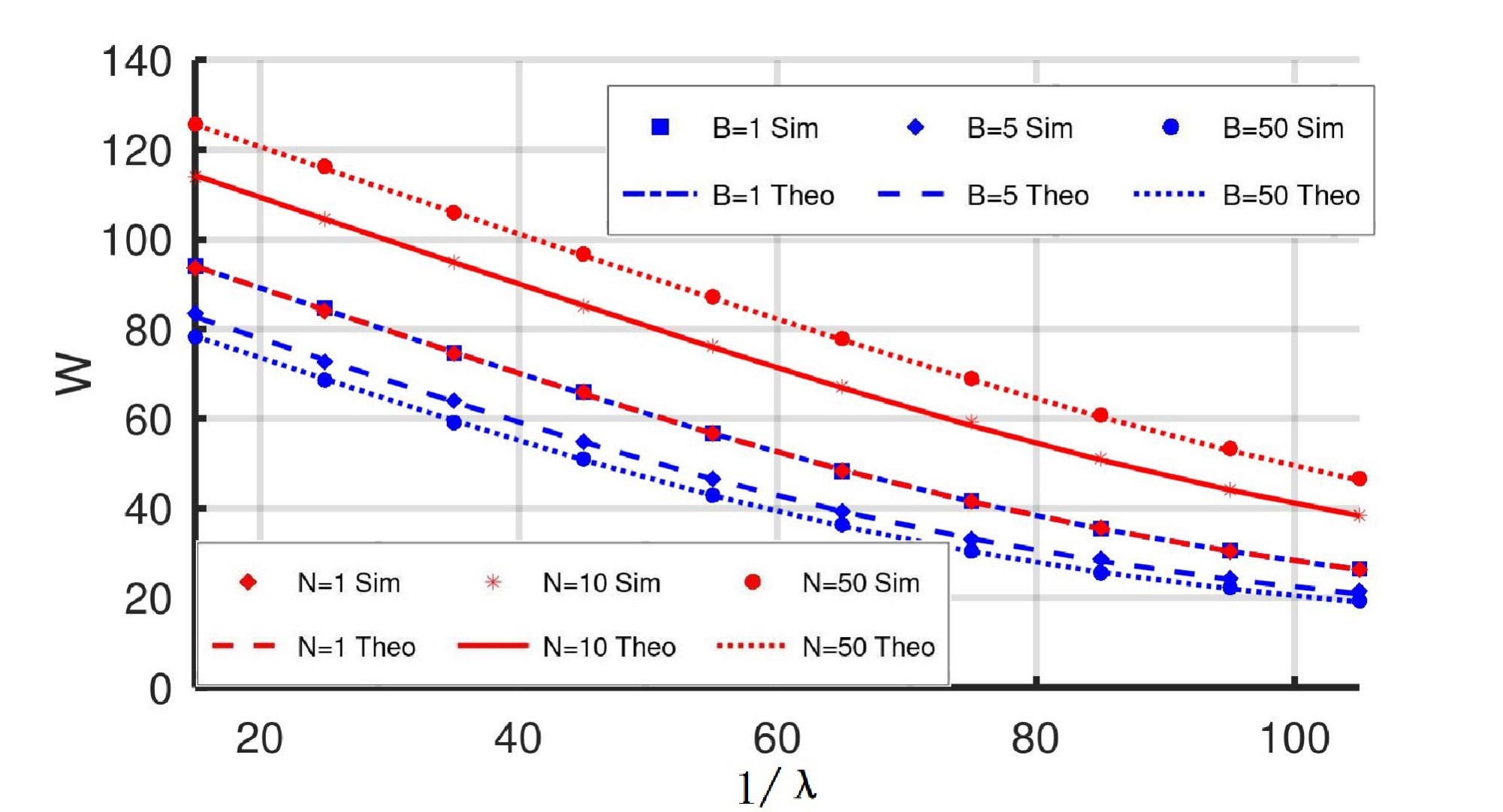}
\caption{Different B and N: Theoretical(from equation) and simulation(from code) results on W.}
\label{fig_11}
\end{figure}

\begin{figure}[ht]
\centering
\includegraphics[width=3.5in]{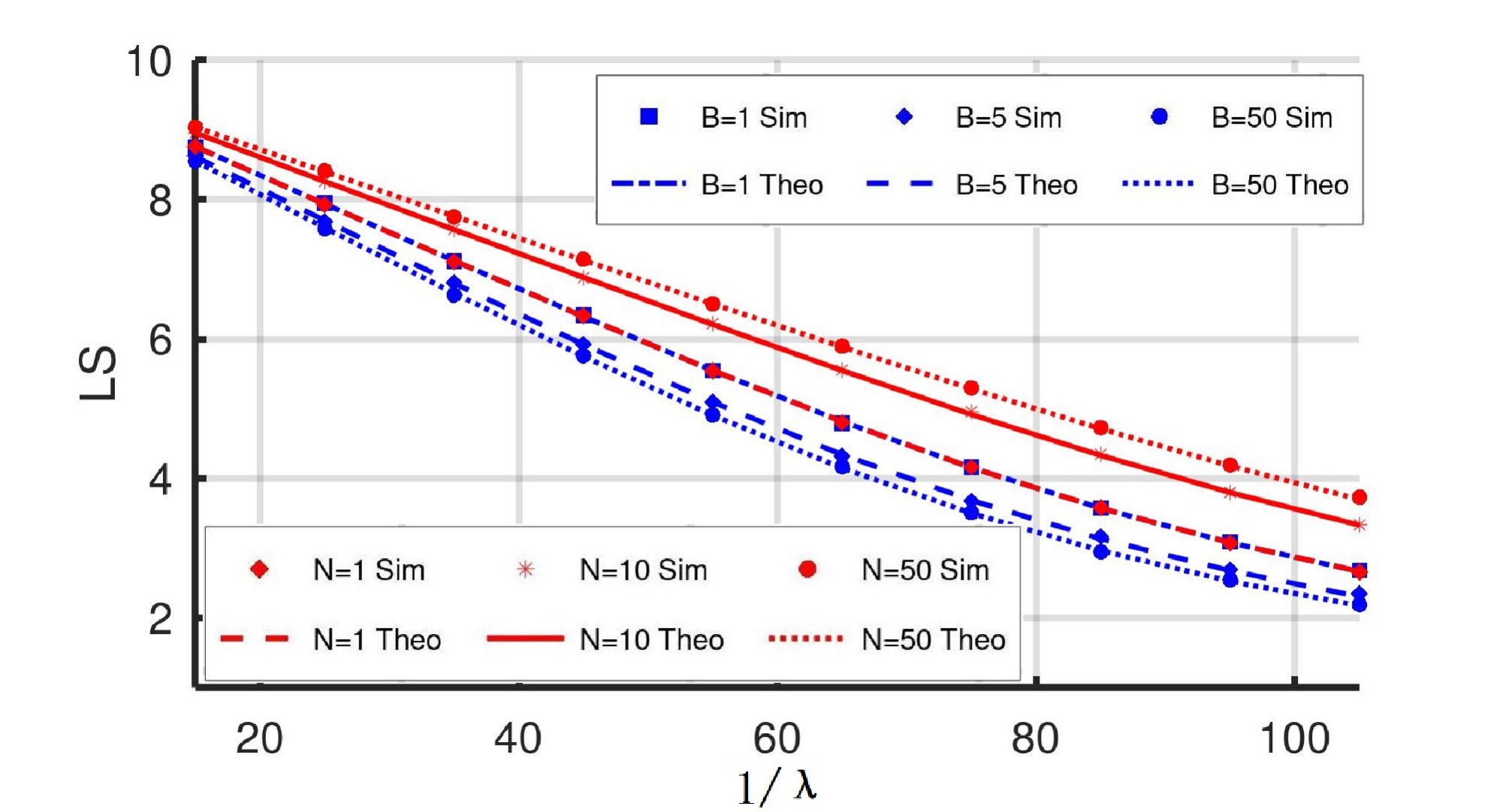}
\caption{Different B and N: Theoretical(from equation) and simulation(from code) results on LS.}
\label{fig_12}
\end{figure}

Fig. \ref{fig_10}, Fig. \ref{fig_11}, Fig. \ref{fig_12} depict WS, W, and LS in different B and N settings. It can be seen that the analysis results of the model and simulation results match each other. In addition to these metrics decreasing as transaction density decreases, it can also be seen that the spacing between lines decreases as B increases. And once the bandwidth increases to a certain extent, neither the lowering of these metrics nor the optimization of the system are significantly affected by it. As N reduces, these metrics also reduce at the same $1/\lambda$. It simply demonstrates that channel quality with low noise can reduce database processing congestion. 

\end{document}